\documentclass{article}

\usepackage{PRIMEarxiv}

\usepackage[utf8]{inputenc} 
\usepackage[T1]{fontenc}    
\usepackage{hyperref}       
\usepackage{url}            
\usepackage{booktabs}       
\usepackage{amsfonts}       
\usepackage{nicefrac}       
\usepackage{microtype}      
\usepackage{lipsum}
\usepackage{fancyhdr}       
\usepackage{graphicx}       
\graphicspath{{media/}}     
\usepackage{amsmath}
\usepackage{subfigure}

\title{Integrated Detection and Tracking Based on Radar Range-Doppler Feature
\thanks{This work was supported in part by the National Natural Science Foundation of China under Grant 61771110 and U19B2017, in part by the Chang Jiang Scholars Program and the 111 Project No. B17008.}
\thanks{Authors’ addresses: University of Electronic Science and Technology of China, School of Information and Communication Engineering, Chengdu, China.}
\thanks{Email address: \href{kussoyi@gmail.com}{kussoyi@gmail.com} (Corresponding author: Wei~Yi)}
}

\author{
  Chenyu Zhang, Yuanhang Wu, Xiaoxi Ma and Wei Yi \\
  University of Electronic Science and Technology of China \\
  Chengdu, China\\
  \texttt{cyzhang57@outlook.com; \{wuyuanhang1677, hazwzmxx\}@163.com; kussoyi@gmail.com} \\
}

\begin{document}
\maketitle

\begin{abstract}
Detection and tracking are the basic tasks of radar systems.  
	Current joint detection tracking methods, which focus on dynamically adjusting detection thresholds from tracking results, still present challenges in fully utilizing the potential of radar signals.
	These are mainly reflected in the limited capacity of the constant false-alarm rate model to accurately represent information, the insufficient depiction of complex scenes, and the limited information acquired by the tracker.
	We introduce the Integrated Detection and Tracking based on radar feature (InDT) method, 
	which comprises a network architecture for radar signal detection and a  tracker that leverages detection assistance. The InDT detector extracts feature information from each Range-Doppler (RD) matrix and then returns the target position through the feature enhancement module and the detection head. The InDT tracker adaptively updates the measurement noise covariance of the Kalman filter based on detection confidence.
	The similarity of target RD features is measured by cosine distance, which enhances the data association process by combining location and feature information.
	Finally, the efficacy of the proposed method was validated through testing on both simulated data and publicly available datasets.
\end{abstract}


\section{Introduction}
Detection 
and tracking are the basic tasks of a radar system. The classic processing chain:  targets are first detected from the echo data based on signal energy thresholds, then the tracker interrogates the detector to obtain state information of the measurements above threshold, and finally the number and state of targets are estimated from measurement points.
The constant false alarm rate (CFAR) \cite{cfar} and the Kalman filter (KF) \cite{Kalman} have been proved to be effective algorithms, and a series of subsequent algorithms \cite{cacfar,phd,pmbm} have been generated, which are applied to various popular frameworks.
\begin{figure}[htp]
	\centering
	\includegraphics[scale=0.57]{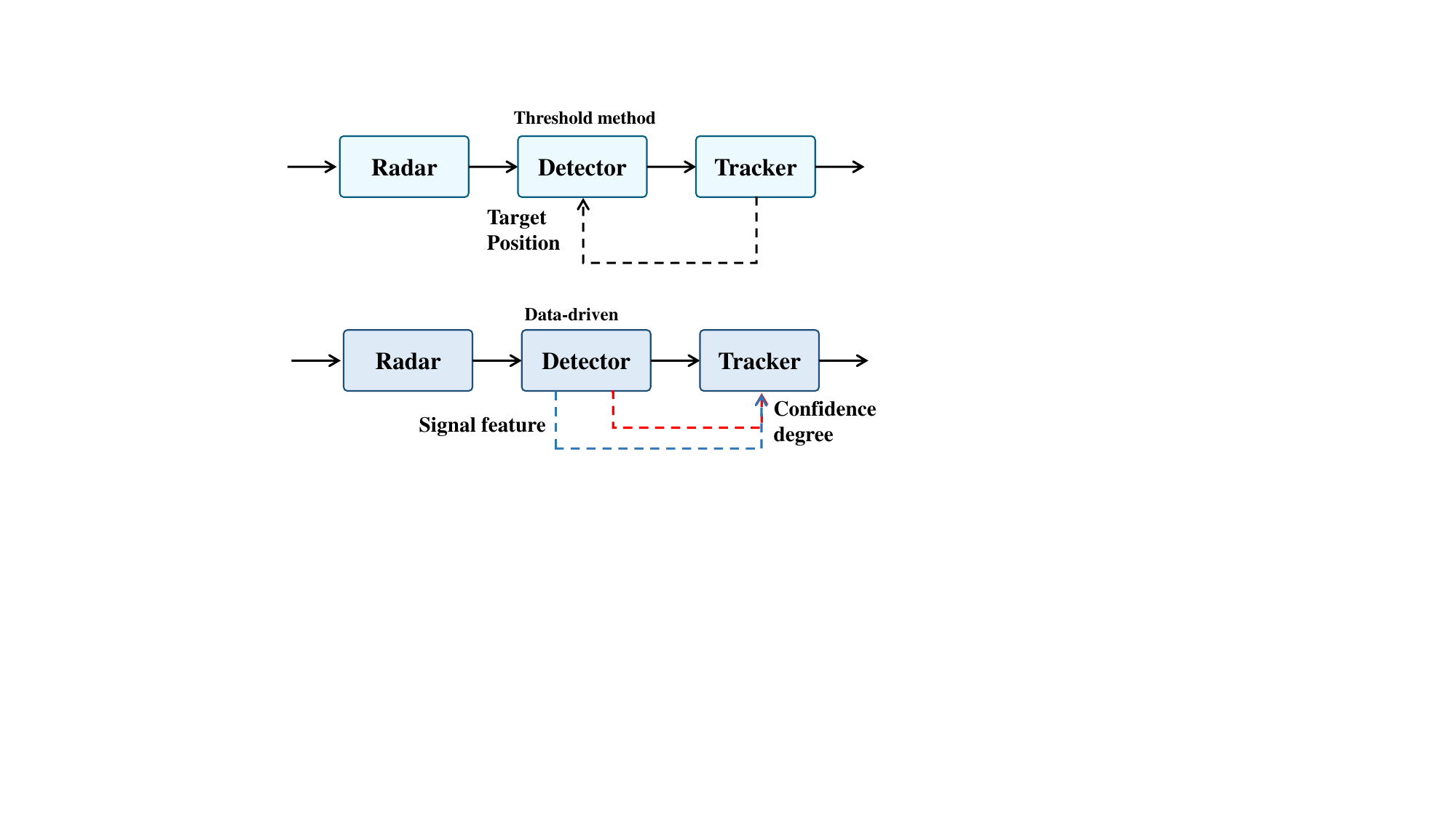}
	\centering
	\caption{Conceptual comparison of the integrated detection and tracking methods.}
	\label{f11}
\end{figure}

In light of the information flow relationship between detection and tracking tasks, several studies have explored joint detection and tracking algorithms \cite{willett2001integration,zeng2012offline,yan2015joint,guan2020joint}. The fundamental concept behind these algorithms is to enhance detection performance by modifying detector thresholds using the predicted measurement position and innovations covariance supplied by the tracker. 
These works let detection and tracking work jointly in an attempt to improve the utilization  of information by the model. 
However, the ability to express information using only traditional model-driven approaches is limited.
Considering the detection task as a complex nonlinear transformation, the threshold decision approach is relatively coarse, resulting in the underutilization of certain signal structure information  \cite{gao2021signal}. 
For example, phase information is lost in the mode-taking operation. 
In addition, the tracker has limited access to radar signal information \cite{gao2023}, making it difficult to  associate data in scenarios with high false alarm rates.
Another approach is track-before-detect (TBD) \cite{tbd1,tbd2}, which utilizes non-coherent integration of multi-frame observation data in the temporal dimension to address the weak target detection problem. The integration process relies on the accuracy of the predefined models.

Recently, data-driven approaches have been shown as a promising alternative to traditional methods \cite{dengjie,wo}. Deep learning can train a network to extract the feature information contained in radar echo signals so as to predict the location of targets in complex scenes \cite{guangxue1,guangxue2,guangxue3}. Under the premise of ensuring accuracy, reasoning speed has obvious advantages. In addition,  MT3 has achieved the same performance as  model-based state-of-the-art (SOTA) algorithm Poisson multi-bernoulli mixture filter (PMBM) \cite{pmbm} in the field of radar multi-target tracking \cite{mt3}. In essence, these data-driven approaches are processed by extracting high-dimensional features of targets, noise, and clutter.

\textit{Is it possible to make better use of signal data by being data-driven, to improve the information utilization of the integrated detection and tracking model.}
We notice the problem of multi-object tracking in the field of computer vision. An accurate and stable framework is "Tracking by detection"\cite{sort,deepsort}, which tracks based on bounding boxes of detection and their appearance feature. The top-performing method (StrongSORT \cite{strongsort}) has achieved SOTA accuracy. 
Inspired by this, we propose an integrated detection and tracking model based on radar signals. And our motivation stems from the development of radar methods and the completely different challenges of high false alarms, sparse information, and so on. The information flow relationship between the detection task and the tracking task has been better utilized as shown in Fig.~\ref{f11}.
\begin{figure*}[!th]
	\centering
        \includegraphics[scale=0.672]{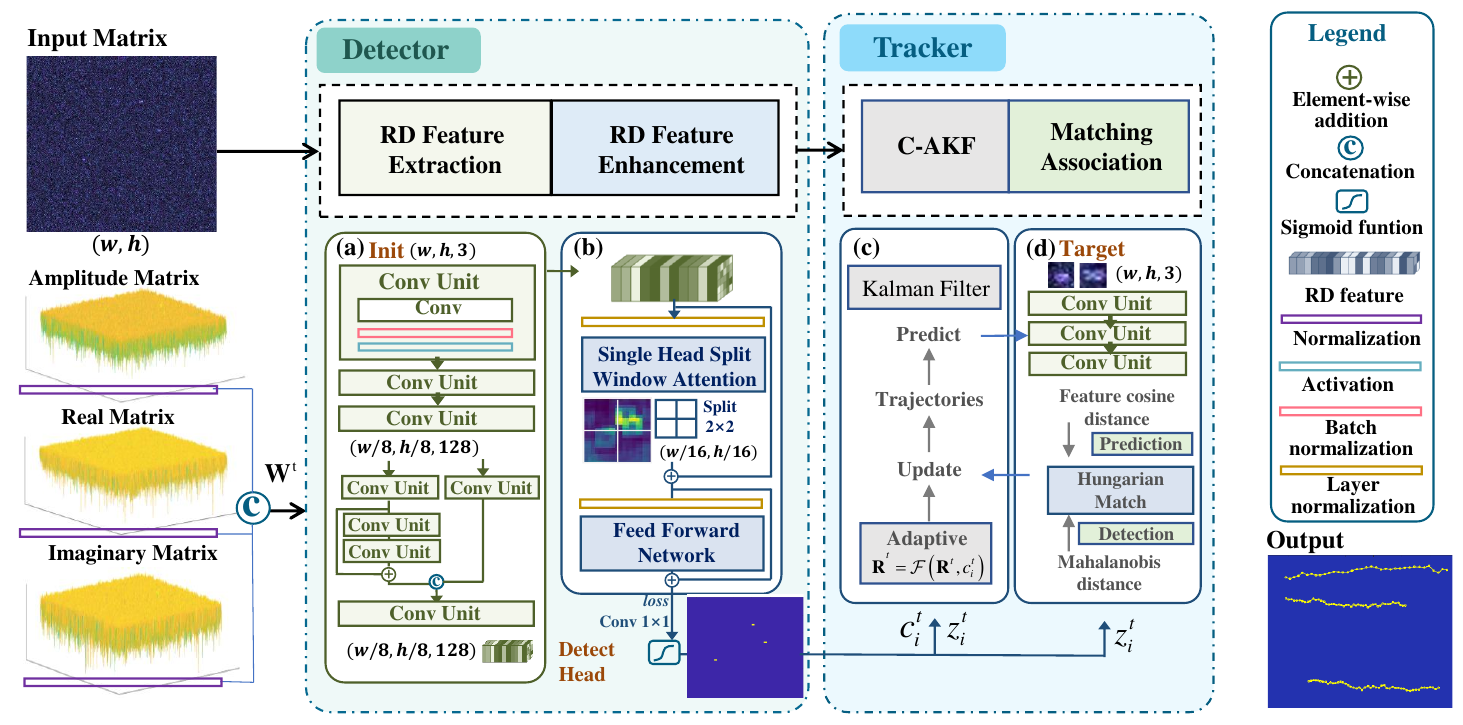}
	\centering
	\caption{InDT consists of 4 main components: (1) A feature encoder that extracts the feature vector of each signal matrix. (2) A feature enhancement layer based on global correlations in the transformer encoded signal matrix. (3) Based on the detection confidence, the measurement covariance of Kalman filter is adjusted, and the prediction and update are made. (4) Matching association, the similarity of RD features is measured by cosine distance, which enhances the data association process by combining location and feature information. }
	\label{f2}
\end{figure*}

We retain the key components of the radar signal processing chain: pulse compression, coherent integration \cite{sec2-3}. Raw echoes are processed into radar Range-Doppler (RD) data as model input, incorporating  data-driven approach to the classical processing chain.
First, the threshold discrimination in the traditional method is replaced by a data-driven detection network that extracts pulse RD features using a deep learning algorithm.  Second, following the filter plus data association tracking structure, in order to obtain more a priori information for the tracker, we provide the confidence of the detection results to the filter and the extracted RD feature information to the data association algorithm. Compared to the classical processing flow, our method uses network to extract RD features, which makes fuller use of signal information in detection and tracking. Fig.~\ref{f2} illustrates the overall InDT framework.
Our major contributions are as follows:
\begin{itemize}
	\item We propose a radar detection and tracking integration framework InDT. The tracker obtains the result confidence     and RD feature information of the detection network, to enhances the filter and data association modules, respectively.
	\item We further propose a data-driven detector, which allows us to better utilise the RD information and pass it on to the tracker.  A  three-channel matrix input method is used to improve the way of processing radar signal matrix data.
\end{itemize}

\section{Preliminaries}
\label{Sec2}

In this section, the raw radar echoes are processed.
The radar emits $N$ echoes \cite{sec2-1}
\begin{equation}
	\begin{split}
		\label{CWsignal}
		s(n, t)=A \cdot \operatorname{rect}\left(\frac{t-n T}{T}\right) \exp (j \cdot(2 \pi f_0(t-n T) 
		+\pi k(t-n T)^2)), n = 0, 1, ..., N - 1,
	\end{split}
\end{equation}
where $A$ denotes the amplitude of the radar transmit signal, $T$ denotes the pulse width, $f_0$ denotes the center frequency of the radar, $k = B / T $ denotes the frequency modulation slope, and $B$ denotes the transmit signal bandwidth. 
When the radar transmit echo encounters a target far away from the radar $R_d$, the echo is reflected when the target is moving at a constant speed relative to the radar radial velocity $v_d$. When the antenna receives the reflected echo from the target, the mathematical expression is
\begin{equation}
	\begin{split}
		\label{TX}
		\begin{aligned}
			s_r(n, t) &=K_r \cdot s\left(n, t-\tau_t\right) 
			&=A_r \exp \{ j \cdot[2 \pi f_0\left(t-n T-\tau_t\right) 
			&~~~+\pi k\left(t-n T-\tau_t\right)^2+2 \pi f_d] \},
		\end{aligned}
	\end{split}
\end{equation}	
where $K_r$ denotes the target emission energy attenuation coefficient, $A_r = K_r \cdot A$ denotes the amplitude of the reflected signal received by the antenna, $\tau_t = 2 \left(R_d + v_dt\right) / c$ denotes the time delay of the target reflected echo, and $f_d = 2f_0v_d / c$ denotes the Doppler caused by the target motion frequency.

The echo signal received by the radar can be regarded as the sum of the target signal and the noise and clutter signals.
\begin{equation}
	\label{RXsignal}
	v(t)=\sum_{i=1}^N s_{r, i}(t)+n(t),
\end{equation}
where $N$ denotes the number of targets, $s_{r, i}$ denotes the reflected echo generated by the ith target, $n(t)$ denotes the noise signal, Gaussian white noise is used in this paper. Before pulse compression, we fix the target signal energy and adjust the noise energy by the signal-to-noise ratio (SNR)  as
\begin{equation}
	\label{eq}
	SNR = 10 \cdot log_{10}(\frac{P_S}{P_N}),
\end{equation}
where $P_S$ denotes the average energy of the target signal, and $ P_N $ denotes the average energy of the noise signal.

Finally, pulse compression and coherent integration \cite{sec2-2,sec2-3} are performed on the  echo signal to obtain the RD matrices.

\begin{figure}[th]
	\centering
	\includegraphics[scale=0.39]{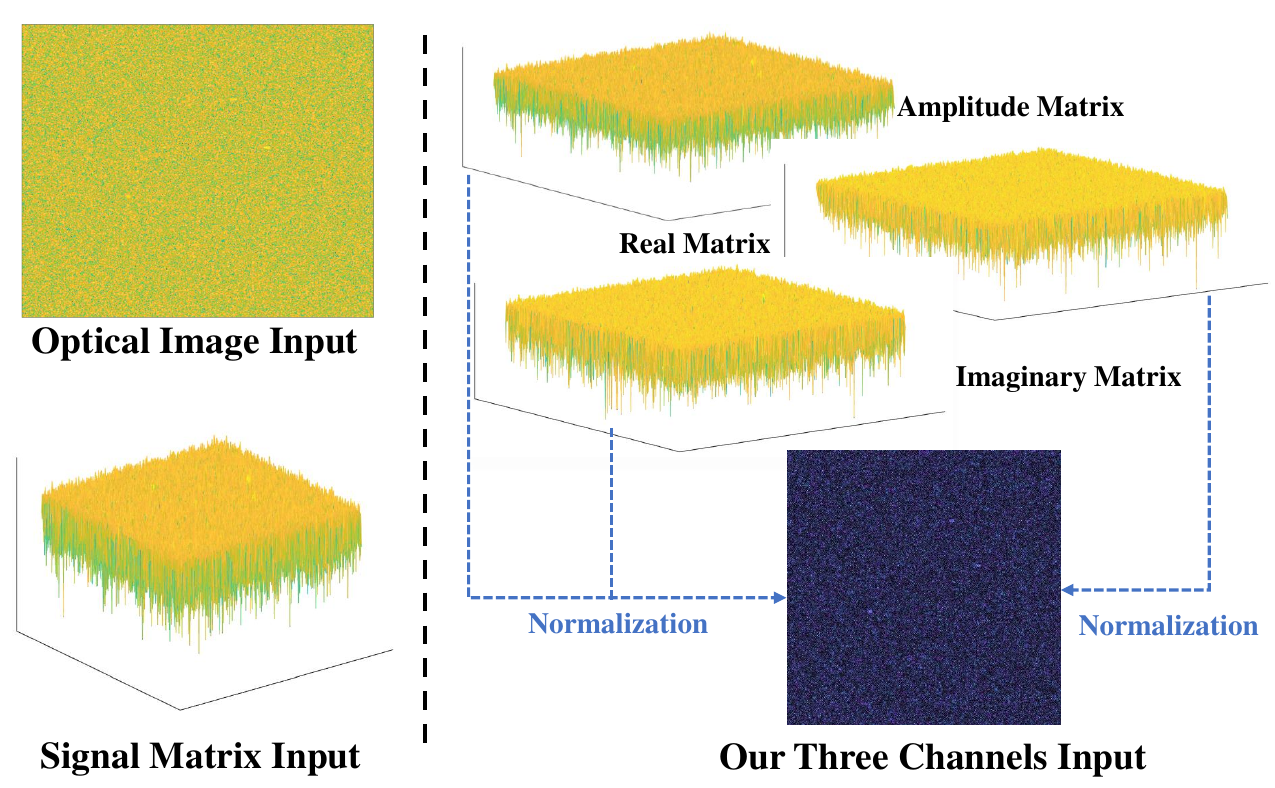}
	\centering
	\caption{A model input method for three-channel complex value signal splicing.}
	\label{f1}
\end{figure}
\section{Approach}
\label{Sec3}

In the following, we first introduce the processing method for three-channel signal matrix input, followed by a detailed description of the components of the InDT algorithm.  An overview of our approach is given in Fig.~\ref{f2}. The detector is divided into two parts: feature extraction and feature enhancement, and through supervised training. 
The tracker is divided into two parts: a filter and data association, which respectively obtain additional assistance from the detection confidence and RD feature information.

Different from converting radar RD maps directly into optical image input \cite{guangxue1,guangxue2} or RD signal energy matrix input \cite{juzhen1,juzhen2,juzhen3}, we redesigned the model input as shown in Fig.~\ref{f1}. The echo signal amplitude matrix, mutually orthogonal real value matrix, and complex value matrix are respectively normalized and spliced into the three channel input matrix as the model input. In this way, the information loss of the transformed optical image and the large computation problem caused by the overall input of the matrix signal are avoided. The normalization operation gathers the sparse information of the signal matrix, and the concate of mutually orthogonal real and complex value matrices can provide more signal details than a single energy channel.

Given a sequence of three channels RD matrices $ \mathbf{W}^{t},$ where  $ t= 1,2,...,N $ indicates the time step of the data. For each RD input $ \mathbf{W}^{t} $, we detect  the state vector $ \mathbf{z}_{i}^{t} \in {\mathbb{R}}^{d_{z}}$ which contains the distance $r_{i}^{t}  $, speed $ v_{i}^{t}  $ and confidence  $ c_{i}^{t} $  of target $ i $ at time-step $ t $. The tracking is conducted based on the detection results. We define the trace result vector of target $ i $  at time-step $ t $ as $ \mathbf{x}_{i}^{t} \in \mathbb{R}^{d_{x}}$ which contains the distance and speed.

\subsection{RD Feature Extraction}
\label{AA}
RD Features are extracted from the input matrices using convolutional networks. The feature encoder network is applied to $ \mathbf{W}^{t} $ and maps the input matrices to dense feature maps at a lower resolution. Our encoder $ f $ outputs features at $ 1/8 $ resolution 
\begin{equation}
	f: \mathbf{W}^{t}\in \mathbb{R}^{R \times D \times 3} \mapsto \mathbf{F}^{t}\in \mathbb{R}^{R / 8 \times D / 8 \times C}, 
\end{equation}
where $ \mathbf{F}^{t} $ stands for the feature obtained by subsampling, we set  feature channel  $ C=128, $ 
$ R $ is the number of sampling points in the range dimension,  $ D $ is the number of pulses, which represent distance and Doppler information, respectively. The convolutional network structure we adopted is shown in Fig.~\ref{f2}. 

To pay more attention to the internal information of each frame signal, we use the instance normalization (IN) layer. And sigmoid weighted linear unit (SiLU) \cite{silu} is used as activation function in \eqref{eq1}
\begin{equation}
	\rm SiLU(x)=\frac{x}{1+e^{-x}}.\label{eq1}
\end{equation}
	
The CSP structure is derived from the cross stage partial network (CSPNet) \cite{csp} combined with the residual network \cite{resnet}. The gradient changes are integrated into the feature map to reduce the computation while maintaining  network accuracy.

\subsection{RD Feature Enhancement}
Since the signal matrix information is sparse. In order to obtain higher-quality detection results,
we further compute the relationship between a resolution cell and the whole RD matrix.
A natural choice is Transfomer \cite{transformer}, which is very suitable for extracting global information with attention mechanism.
We refer to the DETR algorithm \cite{detr} and add two-dimensional sine and cosine position encodings to the extracted feature $ \mathbf{F}_{1} $ to encode the spatial position information.
\begin{equation}
	\widetilde{\mathbf{F}_{t}}=\mathcal{T}(\mathbf{F}_{t}+\mathbf{P},\mathbf{F}_{t}+\mathbf{P}),\label{eq3}
\end{equation}
where $ \mathcal{T} $ is a Transfomer, $ \mathbf{P} $ is the positional encoding, the first input of $ \mathcal{T} $ is query and the second is key and value.

To improve efficiency, we adopt the shifted local window attention strategy from Swin Transformer \cite{swin}. We perform  window multi-head self attention (W-MSA) and multi layer perceptron (MLP) to improve the quality of the initial features $ F_{1}, $ which is mapped as follows
\begin{align}
	\hat{ \mathbf{F}}_{t}&=\rm MSA(LN(\mathbf{F}_{t}))+\mathbf{F}_{t},\\
	\widetilde {\mathbf{F}_{t}}&=\rm MLP(LN(\hat{ \mathbf{F}}_{t}))+\hat{ \mathbf{F}}_{t}, \\
	\mathbf{z}_{i}^{t}&=\rm Head(\rm SPPF(\widetilde {\mathbf{F}_{t}})),
\end{align}
where LN is a layer normalization layer \cite{ln}. 

The spatial pyramid pooling (SPP) realizes the fusion of local and global features, and enriches the expression ability of features \cite{spp}.   The SPPF from the YOLOv5 enhances the computational efficiency by connecting the pooling layers in series.  It is concatenated of four branches: three maximum pooling operations and a shortcut from the input $ \widetilde {\mathbf{F}_{t}} $ as shown in Fig.~\ref{f2}.    Through this free improvement trick, the details of different scales of the target signal are enhanced, and the problem of large changes in the size of the sampling signal matrix is well handled.
Finally, the enhanced signal features are decoded by the detection head regression target position and results' confidence, referring to the detection head of the YOLO series \cite{yolov7}. 

\subsection{Supervision}
We supervised our network between the predicted state and the ground truth state. Given ground truth state vecotor of target $ i $
at time-step $ t $ as $ \mathbf{y}_{i}^{t} \in \mathbb{R}^{d_{y}} $. The loss $ \mathcal{L} $ is  designed to be composed of position loss $ \mathcal{L}_{p} $ and confidence loss $ \mathcal{L}_{c} $ as follows
\begin{align}
	\mathcal{L}&= \sum_{i} \sum_{t}( \lambda_{1}\mathcal{L}_{p} + \lambda_{2}\mathcal{L}_{c}),\\
	\mathcal{L}_{p}& = \left(\mathbf{y}_{i}^{t} -\mathbf{z}_{i}^{t} \right)^{2},  \\
	\mathcal{L}_{c}& =  {o_{i}^{t}}  \cdot \log \left( c_{i}^{t} \right) + \left( {1 - o_{i}^{t}} \right) \cdot \log \left( {1 - c_{i}^{t}} \right), 
\end{align}
where  $o_{i}^{t}$ represents the model predicts the probability that the sample is positive. We set $ \lambda_{1} = 0.7, \lambda_{2}=0.3 $ in our experiments.

\subsection{Adaptive Kalman Filtering Based on Detection Confidence (C-AKF)}
We used KF \cite{Kalman} to predict and update the state of the target. Under the classical KF framework, combined with the confidence of the detection results, the covariance  of the measurement noise $ \mathbf{R} $ was adaptively updated.

The prediction confidence of the detector is directly related to the SNR of the scene.  The lower the detection confidence is, the greater the input signal noise is, and the covariance  $ \mathbf{R} $ should be increased. We define a mapping $ \mathcal{F} $ to reflect the inverse relationship between detection confidence $ c_{i}^{t} $ and  $\mathbf{R}_{i}^{t} $, and the adaptive update measurement noise covariance is as follows 

\begin{equation}
	\hat{\mathbf{R}}^{t}=\mathcal{F}\left( \mathbf{R}^{t}, c_{i}^{t}\right).
\end{equation}

In this paper, we use a simple function \eqref{eqk} to verify the above inverse relationship, and the improved C-AKF has better tracking performance than the original KF

\begin{equation}
	\mathcal{F}\left( \mathbf{R}^{t}, c_{i}^{t}\right)=\frac{ n }{2 \sum_{i=1}^{n} c_{i}^{t}}\cdot \mathbf{R}^{t},
	\label{eqk}
\end{equation}
where $ n $ represents the number of detection results in time-step $ t $. Finding a more appropriate function is an open question.

\subsection{Data Association}
In traditional tracking tasks, data association is performed solely based on location information.  In this section, we enhance the matching process by incorporating  RD feature information in addition to location information.  The similarity of features is measured using cosine distance, thereby improving the accuracy of data association.

The correlation of position information is measured by the Mahalanoulian distance between the state predicted by the Kalman filter and the newly arrived measurement. We define $ d^{(1)} $ to represent the motion matching degree between the $ i $-th detection result and the $ j $-th target trajectory.
\begin{equation}
	d^{(1)}(i, j)=\left(\boldsymbol{z}_{i}-\boldsymbol{x}_{j}\right)^{\mathrm{T}} \boldsymbol{S}_{j}^{-1}\left(\boldsymbol{z}_{i}-\boldsymbol{x}_{j}\right),
\end{equation}
where $ \boldsymbol{S}_{j} $ is the covariance matrix of the measurement space at the current moment, which is predicted by the KF.

For each detected target, we use a simple convolution layer to extract its RD  features, and obtain the feature vector with the number of channels of 64, which is regularized to the hypersphere of a unit sphere (modulus length 1).
The  RD  network here are obtained by off-line pre-training in advance. With the arrival of the new detection measurement, the minimum cosine distance between the feature $ \mathbf{f}_{j} $ of the $ j $-th trajectory and the detection feature $ \mathbf{f}_{i} $ of the $ i $-th detection results is
\begin{align}
	d^{(2)}(i, j)&=\min \left\{1-\mathbf{f}_{i}^{T}\mathbf{f}_{j} \right\}, \\
	\mathbf{f}_{j}^{t}&=\alpha \mathbf{f}_{j}^{t-1}+(1-\alpha) \mathbf{f}_{i}^{t},\label{eq10}
\end{align}
for each track, an exponential moving average (EMA) \cite{strongsort} feature update strategy was adopted in \eqref{eq10}, and only the most recent time step feature was saved. We set $ \alpha=0.7. $

Position information and feature information can complement each other. On the one hand, Mahalanobis distance provides position correlation based on motion prediction; on the other hand, when motion information has weak discriminative power, cosine distance will consider the feature information of target points. We use a weighted sum to combine the two distances in \eqref{eq11}.
\begin{equation}
	s_{i, j}=\mu d^{(1)}(i, j)+(1-\mu) d^{(2)}(i, j),
	\label{eq11}
\end{equation}
where we set $ \mu=0.3 $, $ s_{i, j} $ formed the cost matrix $ \mathbb{C} $, modeled the data association problem as the minimum cost assignment problem, and solved according to the Hungarian algorithm. Then the KF is used to update the retained track for online multi-target tracking.

\section{Simulation Results}
We evaluate InDT through a series of simulation experiments. In terms of detection performance, the algorithm has a large advantage in detection rate ($ P_{d} $) compared with CA-CFAR and Monte Carlo threshold method under the approximate false alarm rate ($ P_{fa} $). In terms of tracking performance, we process radar range tracking in a multi-target environment with low SNR, and the InDT tracking accuracy outperforms the PMBM.

\begin{small}
	\begin{table}[htbp]
		\caption{Simulated Radar Parameters}
		\begin{center}
			\renewcommand{\arraystretch}{1.01}\setlength{\tabcolsep}{3mm}{\begin{tabular}{c  l  c}
					\toprule
					\textbf{Parameter} & \textbf{Definition} & \textbf{Value} \\
					\hline
					$ f_0 $ & Carrier frequency & $ 77 $GHz \\
					\rm$ B $ & Sweep Bandwidth & $ 561.96 $MHz \\
					\rm$ L $ & Num of chirps in one frame & $ 512 $ \\
					\rm$ M $ & Num of samples of one chirp & $ 512 $ \\
					\bottomrule
			\end{tabular}}
			\label{tb11}
		\end{center}
	\end{table}
\end{small}
\begin{figure}[htp]
	\centering
	\includegraphics[scale=0.36]{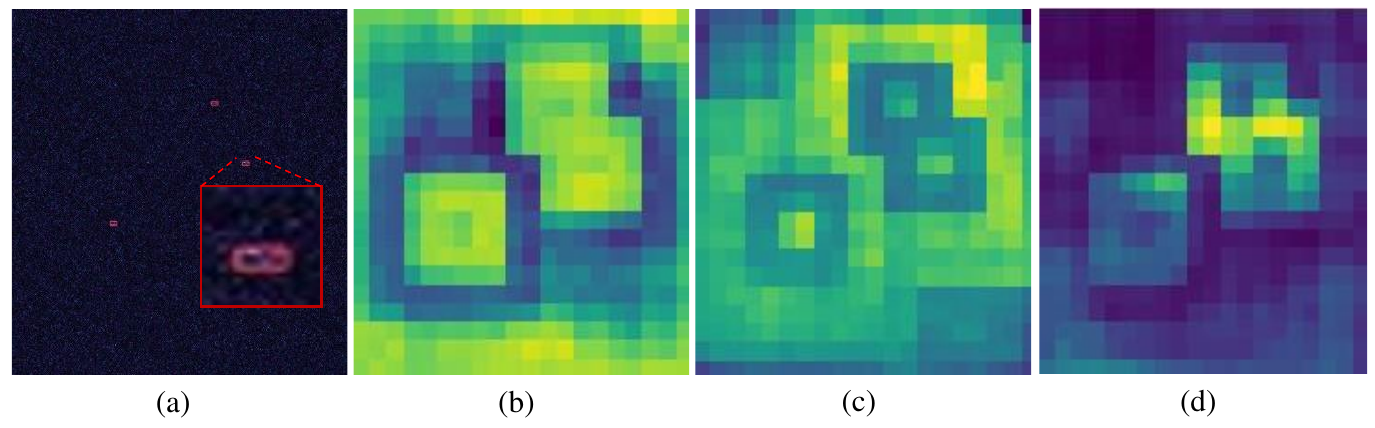}
	\centering
	\caption{Visualization of detection: (a) SNR=$ -20 $dB signal matrix detection result. (b) Amplitude channel feature map. (c) Real-valued channel feature map. (d) Complex-valued channel feature.}
	\label{d1}
\end{figure}
\textbf{Data generation.} The simulation data follows the radar processing procedure in Section \ref{Sec2} to generate the raw radar echo data with different SNR. We use the Linear Frequency Modulated Continuous Wave (LFMCW) signal and the parameter settings are shown in Table \ref{tb11}.
The training set and test set are divided according to $ 7:3 $.

\textbf{Metrics.}   The detection rate ($ P_{d} $) is measured at the same level of false alarm rate ($ P_{fa} $). In the simulation data, we used different SNR scenarios, it should be noted that the SNR here is before signal processing, pulse compression and coherent integration have roughly $ 10\cdot log_{10}(512)\approx27.09\rm{dB} $ gain. We adopt the commonly used metric in multi-object tracking, i.e., the optimal subpattern assignment (OSPA) \cite{ospa}, which is the average distance between the prediction and ground truth sets of objects with optimal matching and cardinality penalty.

\textbf{Experimental details.} 
We implement our detector in PyTorch. Our convolutional backbone network is identical to YOLOv5 model, except that our final feature dimension is $ 256 $. We further stack $ 2 $ Swin Transformer blocks to enhance the representation of the radar RD features. 
To enhance the robustness of the detection model under different SNR environments, we performed data enhancement before training by randomly adding Gaussian noise to the radar signal matrix.  The model was trained using Adam optimizer  and initial learning rate $ 0.01 $.  The model was trained for $ 100 $ epochs, and turn off data enhancement $ 3 $ epochs in advance.

\subsection{Detection Scenarios with Different SNR}
We test the trained InDT detector under different SNR scenarios. Fig.~\ref{d1}(a) depicts the detection results of a single signal matrix when SNR= $ -20\rm{dB} $  (About $7\rm{dB} $  after signal processing). 
Fig. \ref{d1}(b)-(d) shows the visualisation of the RD feature $ \mathbf{F}_{t} $ in \eqref{eq3}, demonstrating the features extracted and recognised by the neural network on different channels.

\begin{figure}[htbp]
	\centering
	\subfigure[] {
		\centering
		\includegraphics[width=2.75in]{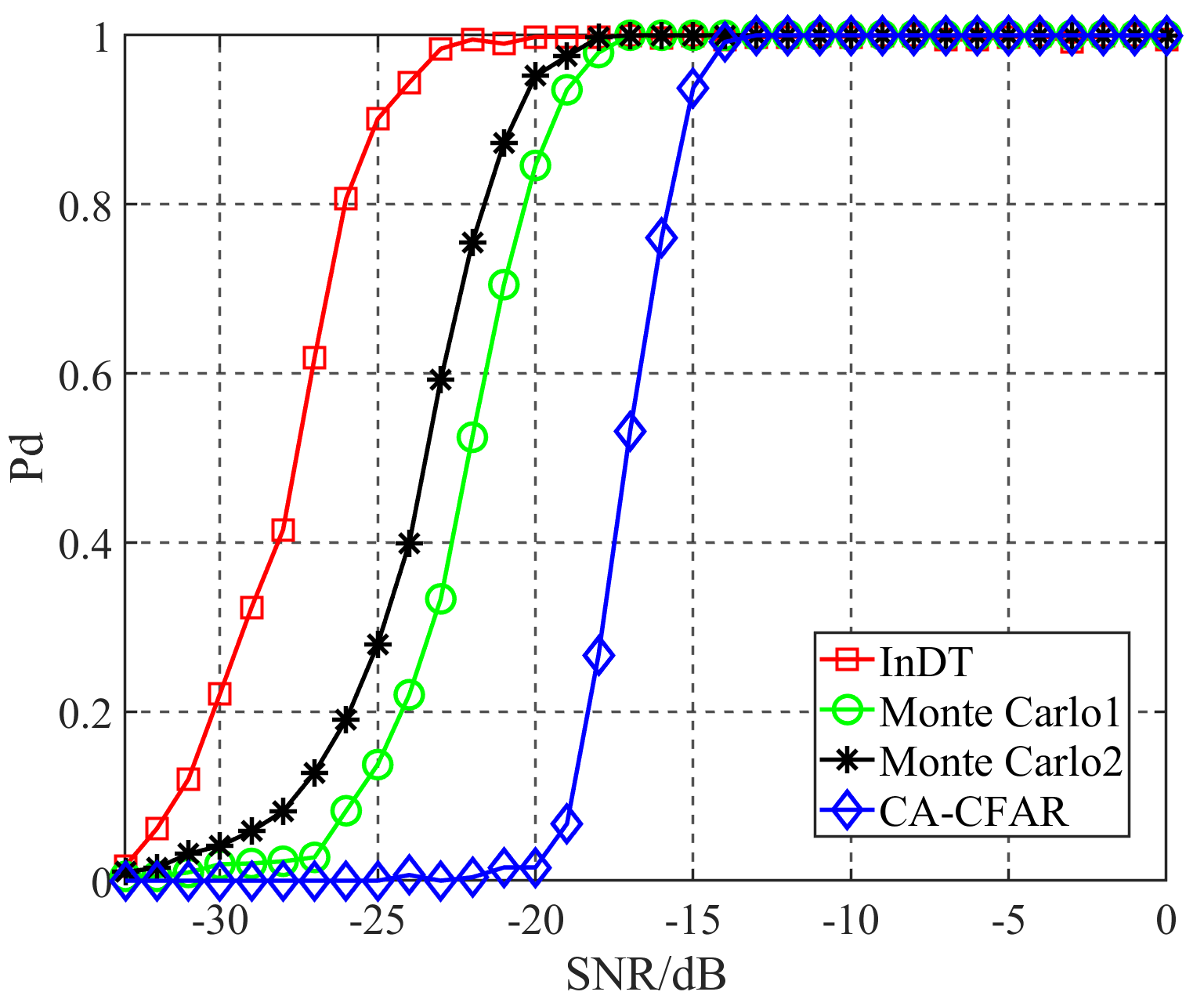}  
	}
	\centering
	\subfigure[] {
		\centering
		\includegraphics[width=2.75in]{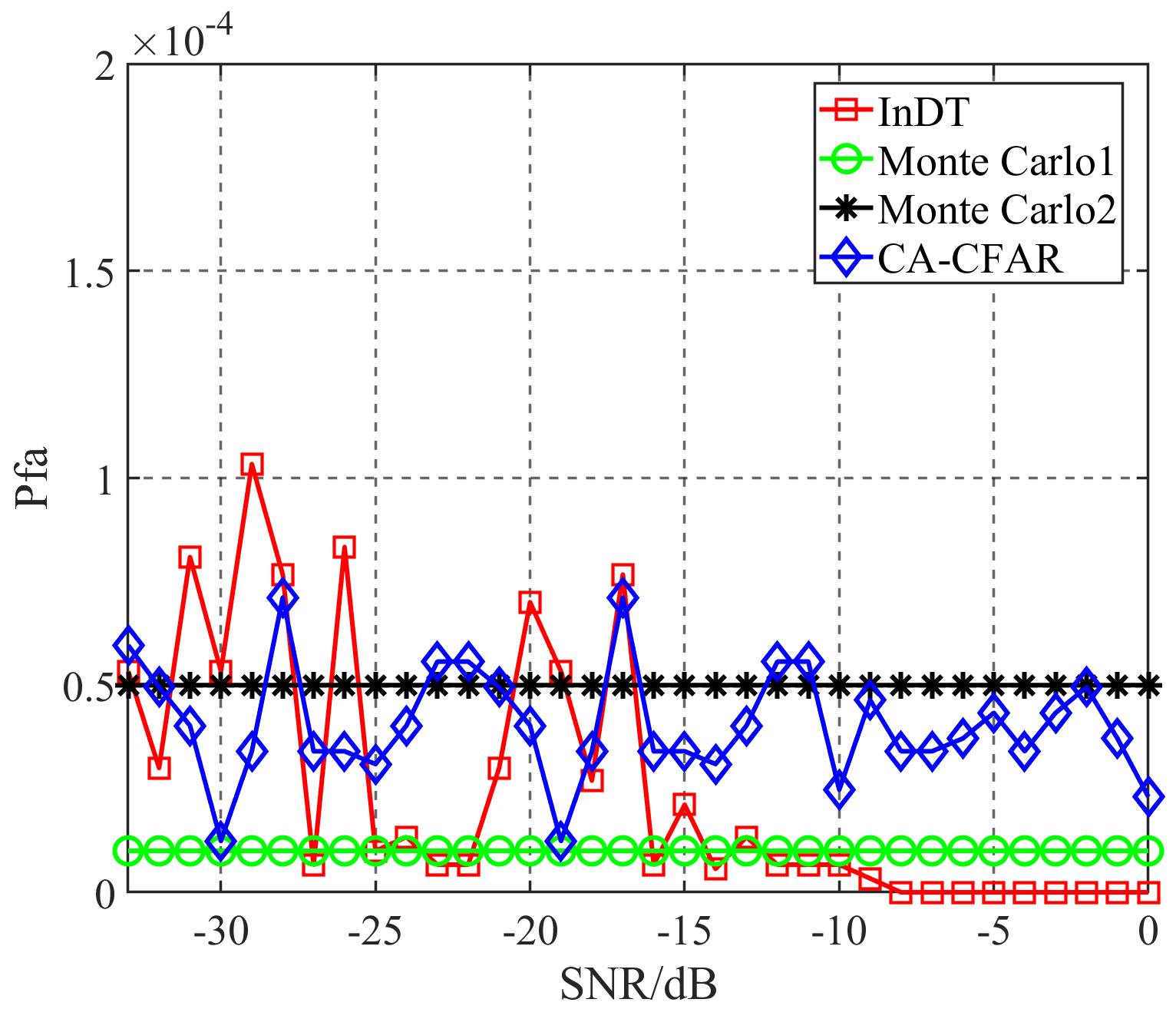}
	}
	\centering
	\caption{Detection performance: (a) The $ P_{d} $  are
		plotted against SNR. (b) The corresponding $ P_{fa} $  are
		plotted against SNR.}
	\label{r1}
\end{figure}
The Monte Carlo threshold method approximates the theoretical upper limit of the threshold detection algorithm. In the simulation experiment, we set a fixed false alarm  value to limit the number of false alarms that can be detected. All clutter energies are sorted, and the clutter energy at the junction is calculated as the detection threshold. Since the traditional joint detection tracking algorithm dynamically adjusts the energy threshold for detection through the tracking results \cite{willett2001integration,guan2020joint}, the Monte Carlo's computational method can be seen as a stronger benchmark model.

\begin{figure}[htp]
	\centering
	\includegraphics[scale=0.27]{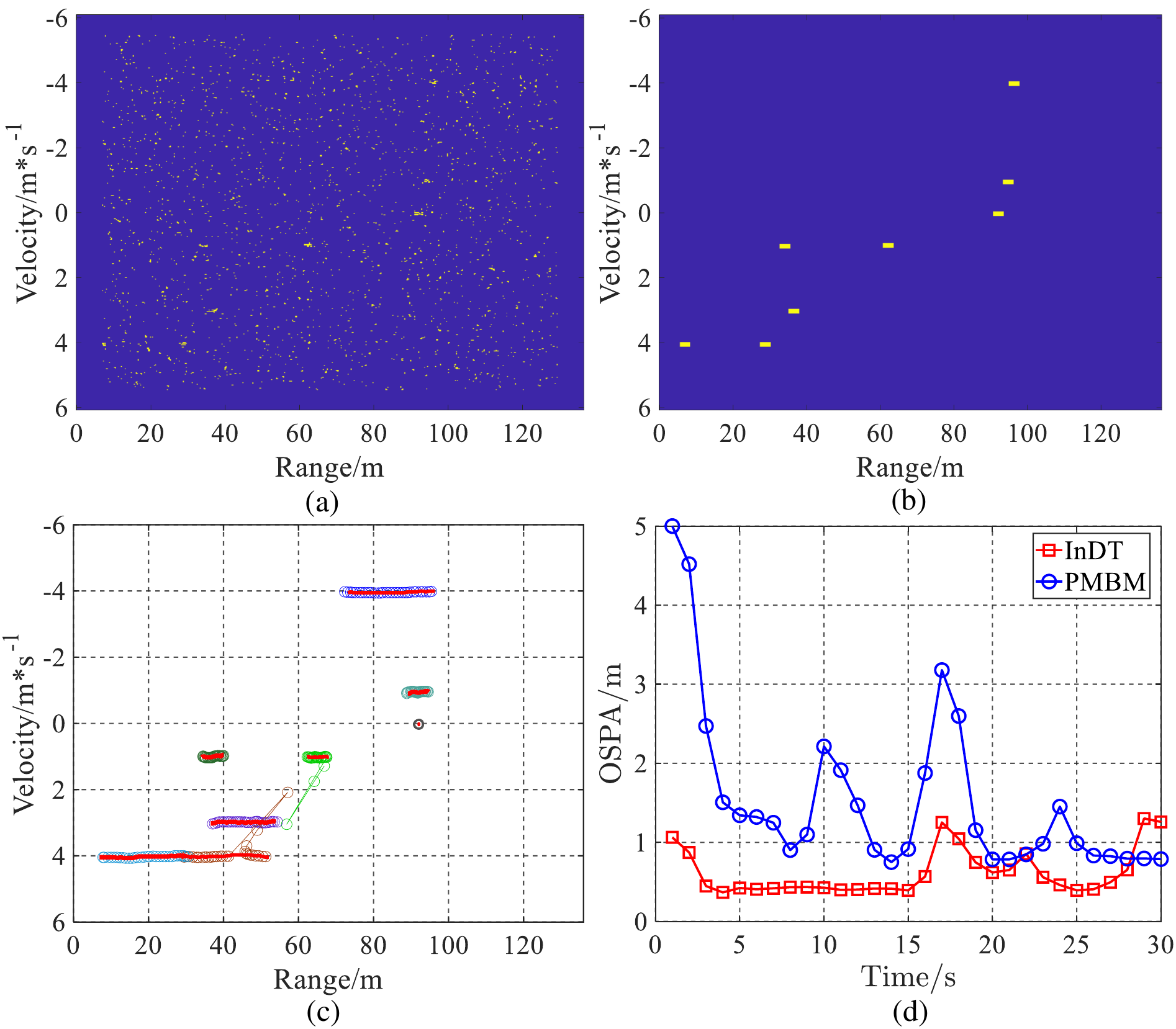}
	\centering
	\caption{Multi-target tracking scenario with low SNR. (a) Single frame measurement. (b) Single frame ground truth. (c) Ground truth and InDT tracking trajectories are shown in red and other colors, respectively. (d)  The OSPAs of range (m) are plotted against time for InDT and PMBM. }
	\label{r2}
\end{figure}
Fig.~\ref{r1} presents the detection results.
We compare two methods Monte Carlo 1 and Monte Carlo 2 with fixed false alarm rates $ 10^{-5} $ and $ 5\cdot 10^{-5} $ respectively. 
At the same level $ P_{fa} $, 
InDT achieves a gain of about $ 10 $dB and $ 4 $dB over CA-CFAR and Monte Carlo 2 method respectively.
The superior detection performance of the InDT detector stems from the excellent feature extraction ability of the model and the three channel RD matrix input,
which preserves more signal information by superimposing the channels after  normalization.


\subsection{Multi-target Tracking Scenario with Low SNR}


We evaluate the impact of feature information on data association and the filtering performance of C-AKF in a low-SNR multi-target environment.
To ensure fair comparison in tracking and to eliminate the influence of different detection performance,
we set the SNR=$ -20$dB (About $7$dB after signal processing), where the CFAR algorithm produces a large number of false alarms while maintaining the target detection. We adjust the detection confidence of InDT to match the same false alarm scenario as the PMBM algorithm.
The tracking scenario we use is depicted in Fig.~\ref{r2}(a)-(b), where each time step has 7 to 10 targets and more than 3000 clutter points.

Convert Doppler to velocity to track the distance dimension. The trajectory follows the constant velocity model and lasts for $ 30 $ frames.   The PMBM algorithm has a fixed initial point.  The state transition and measurement matrices are
\begin{align}
	\mathbf{F}=\left[\begin{array}{cc}
		1 & \Delta t \\
		0 & 1
	\end{array}\right],\quad
	\mathbf{H}=\left[\begin{array}{cc}
		1 & 0 \\
		0 & 1
	\end{array}\right],
	\label{25} 	
\end{align}
where $ \Delta t =0.2s $. The covariance between the process noise and the measurement noise is constant and given by
\begin{equation}
	\mathbf{Q}=q_{s}*\left[\begin{array}{cc}   
		\Delta t^{3}/3 & \Delta t^{2}/2 \\
		\Delta t^{2}/2 & \Delta t
	\end{array}\right]
	,\quad 
	\mathbf{R}=\left[\begin{array}{cc}
		\sigma_{r}^{2} & 0 \\
		0 & \sigma_{v}^{2}
	\end{array}\right],	
\end{equation}
where  $ q_{s}=0.2 {m^{2}/s^{3}}$ is the process noise intensity, $ \sigma_{r} = 0.6m, \sigma_{v}=0.2 m/s$ are the standard deviations of the measurement noise.
The DBSCAN method is used to cluster the measurement points based on the density and then track them.  Fig. \ref{r2}(d) shows the performance comparison, where we set the OSPA parameters $ c=5 $ and $ p=1 $.
\begin{small}
	\begin{table}[htp]
		\caption{Experimental  Radar Parameters}
		\begin{center}
			\renewcommand{\arraystretch}{1.01}\setlength{\tabcolsep}{3mm}{\begin{tabular}{c  l  c}
					\toprule
					\textbf{Parameter} & \textbf{Definition} & \textbf{Value} \\
					\hline
					$ f_0 $ & Carrier frequency & $ 77 $ GHz \\
					\rm$ B $ & Sweep Bandwidth & $ 670 $ MHz \\
					\rm$ L $ & Num of chirps in one frame & $ 255 $ \\
					\rm$ M $ & Num of samples of one chirp & $ 128 $ \\
					\rm$ f $ & Frame rate & $ 30 $ FPS \\
					\bottomrule
			\end{tabular}}
			\label{tb22}
		\end{center}
	\end{table}
\end{small}
\begin{small}
	\begin{table}[h!]
		\caption{The detection results of the measured dataset.}
		\setlength{\tabcolsep}{16pt}
		\begin{center}
			\begin{small}
				\begin{tabular}{ccc}
					\toprule
					\textbf{Data} & \textbf{CFAR} & \textbf{InDT}
					\\
					\hline
					\text{``2019\_04\_09\_bms1000''} & $ 0.995 $ & $ 0.999 $ \\
					\text{``2019\_04\_09\_cms1000''} & $ 0.881 $ & $ 0.898 $\\
					\text{``2019\_04\_09\_pms1000''} & $ 0.989 $ & $ 0.992 $ \\
					\bottomrule
				\end{tabular}
				\label{tb122}
			\end{small}
		\end{center}
	\end{table}
\end{small}
\begin{figure}[th]
	\centering
	\includegraphics[scale=0.188]{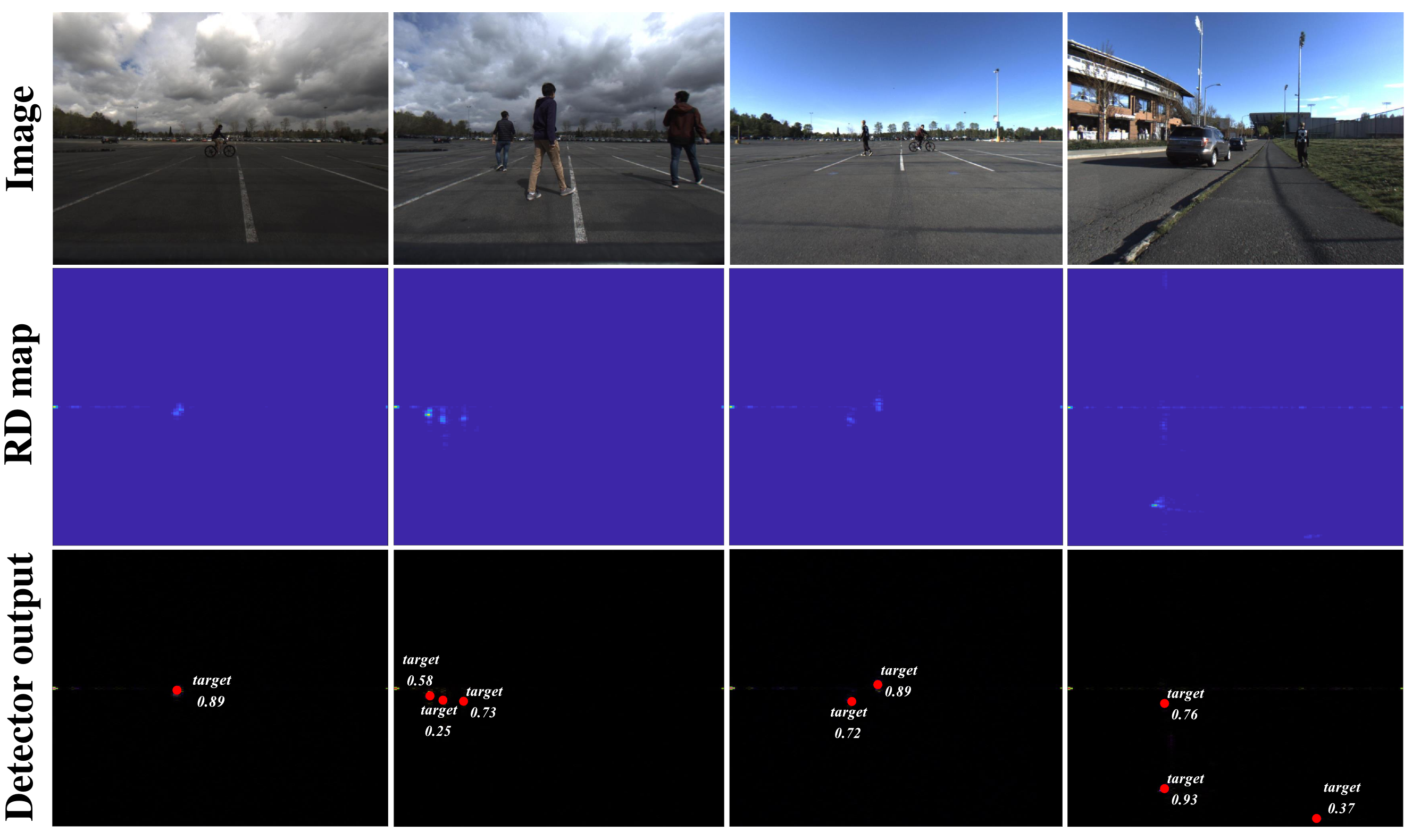}
	\centering
	\caption{Columns 1-3 show the car park scene, and the fourth column shows the roadside scene. For each column, the first row is the synchronized camera image, the second row is the corresponding radar RD plot, and the third row is the visualization of the model detection results.}
	\label{f555}
\end{figure}

Fig. \ref{r2} shows that some points exhibit large fluctuations due to the auxiliary association of feature information. However, the overall OSPA error for InDT is lower than that for PMBM, which confirms the effectiveness of InDT data association and filtering in the noisy multi-target environment. Feature information can assist the tracker in reducing the number of missed targets and false matches, especially in the noisy environment, which benefits the data association. Moreover, C-AKF updates the measurement noise using the detection confidence, which results in lower OSPA for stationary tracking. It outperforms KF with fixed parameters.

\section{Experimental Results}
In this section, the effectiveness of the InDT algorithm is verified on an available public dataset  \cite{data}.  It provides raw millimetre-wave radar echo data for automotive object detection, as well as synchronised camera images and labels. Table \ref{tb22} summarises the radar parameters of this dataset.

The dataset was collected by a vehicle-mounted platform equipped with binocular cameras and radar, which captured various scenes such as car parks and roadsides. A detection model and an unsupervised depth estimation model were applied to the camera images to obtain the position labels of the targets \cite{data}, which we use as the ground truth for distance-dimensional tracking evaluation. We use the CFAR to maintain a high detection rate  to generate radar RD detections, which are manually calibrated and saved as the ground truth required for training and evaluation of the detection task.

\begin{figure}[htbp]
	\centering
	\subfigure[] {
		\centering
		\includegraphics[width=2.3in]{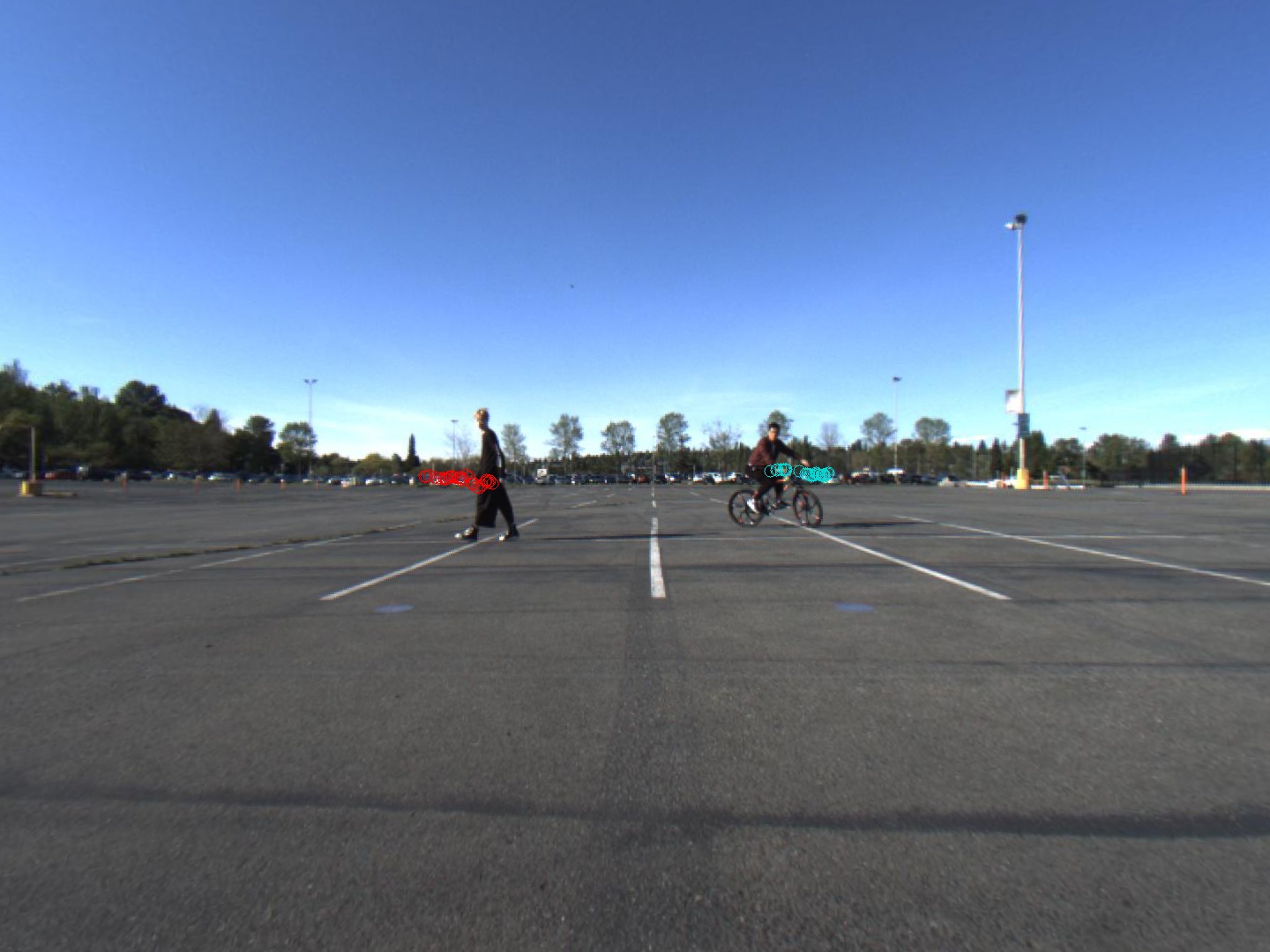}    
	}
	\centering
	\subfigure[] {
		\centering
		\includegraphics[width=2.9in]{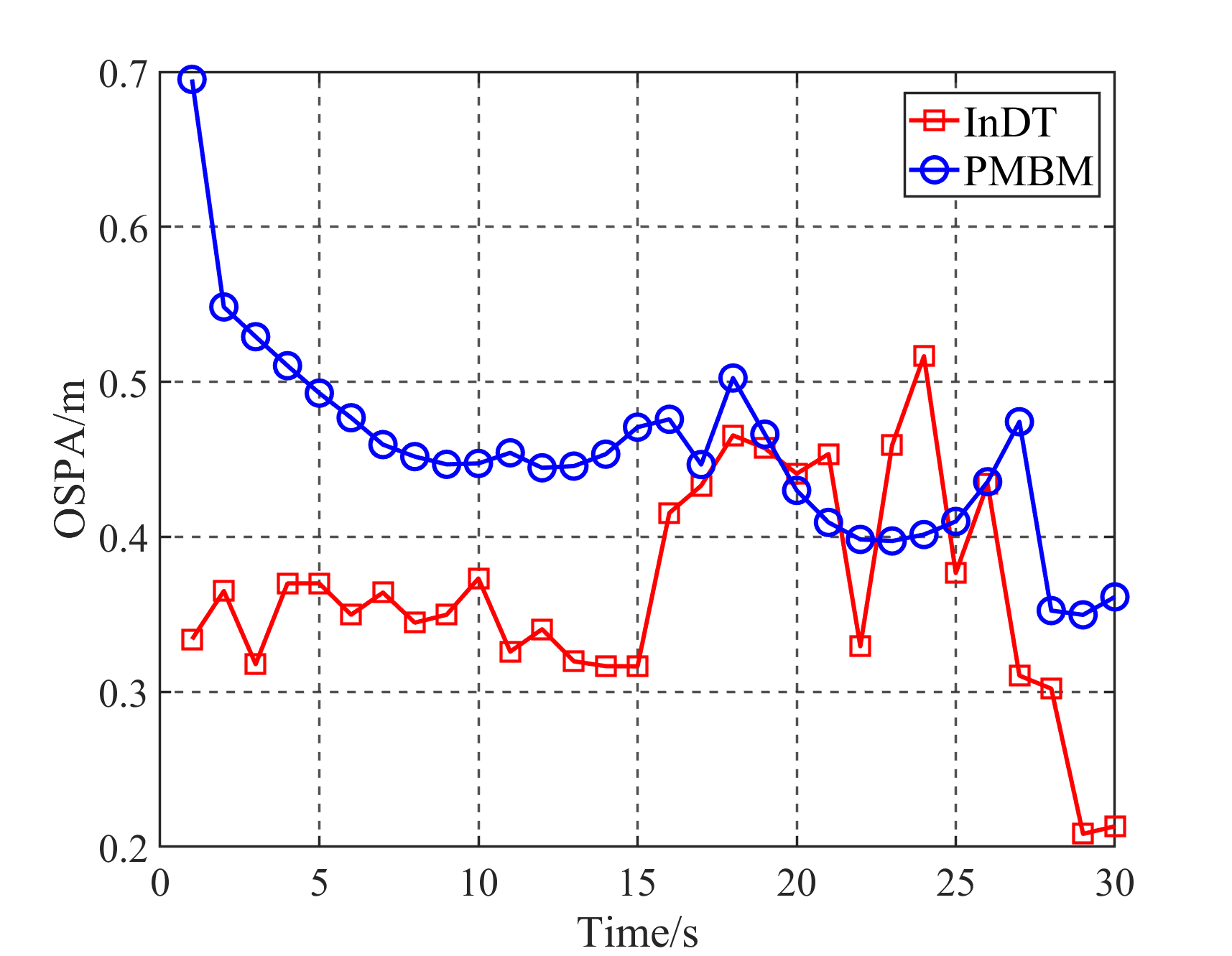}
	}
	\centering
	\caption{(a) Visualisation of target trajectories with synchronised camera images. (b) The OSPAs of range (m) are plotted against time for InDT and PMBM.}
	\label{r5}
\end{figure}

Fig.~\ref{f555} illustrates the detection results for different scenarios. Table \ref{tb122} presents the detection rates for the three datasets in the car park scenario, which have similar backgrounds and a false alarm rate of approximately $ 10^{-4} $. The datasets are: ``2019\_04\_09\_bms1000'', consisting of 897 frames of a bicycle moving along a circular and S-shaped trajectory; ``2019\_04\_09\_cms1000'', consisting of 898 frames of a car moving along a straight line with constant velocity; and ``2019\_04\_09\_pms1000'', consisting of 898 frames of a pedestrian moving along a straight line with constant velocity, followed by a uniform turn.
To handle the large extended targets in the measured data, we apply DBSCAN point trace coalescence to the detection results and use the intersection over union (IoU) with the ground truth shape as the detection criterion. We consider a detection as successful if the IoU is greater than 0.3. Table \ref{tb122} demonstrates that the proposed detector outperforms the CFAR algorithm in terms of detection performance. The lower detection rate for the car target in ``2019\_04\_09\_cms1000'' is due to its larger extension, which imposes a stricter false alarm constraint.

The tracking scene selects ``2019\_04\_30\_pbms002'' data, take 100-130 frames of this data. The real motion trajectories of the two targets are shown in Fig.~\ref{r5}(a). A person and a bicycle turning in different directions at constant velocity.
The extended target DBSCAN clustering neighbourhood search radius is $ 0.5 $ and the minimum number of neighbourhood targets is $ 8 $. In the PMBM algorithm,  the radar scanning period $ \Delta t=1/f\approx0.033s $, $ q_{s}=0.5 m^{2}/s^{3} $, and $ \sigma_{r} = 0.6m, \sigma_{v}=0.5 m/s $. Fig.~\ref{r5}(b) demonstrates the tracking performance and verifies the effectiveness of the proposed algorithm.

\section{Conclusion}

We design the Integrated Detection and Tracking (InDT), a  framework based on radar RD Feature. 
The RD features are extracted by deep learning algorithms, and the detection result confidence and measurement feature similarity are provided to the tracker to make better use of the radar signal information.
In terms of detector performance,  it outperforms the constant false-alarm rate algorithm through excellent signal feature extraction capability. In terms of tracking performance, it is slightly better than PMBM algorithm in multi-target strong noise environment. InDT alsohas the advantages of strong robustness, no prior adjustment of parameters, and fast reasoning speed. In the future, it is possible to continue to build on the signal features to continue to integrate the target classification task in the algorithm.

\bibliographystyle{unsrt}  
\bibliography{main}

\end{document}